\documentclass[%
 preprint,
 amsmath,amssymb,
 aps,
prc,
]{revtex4-1}
\usepackage[dvipsnames,usenames]{xcolor}
\usepackage{graphicx}
\usepackage{dcolumn}
\usepackage{bm}
\usepackage{mathrsfs,natbib}
\usepackage{graphicx,epsf,amssymb,amsbsy,amsfonts,amssymb,amsmath}
\usepackage{slashed}
\usepackage{comment}
\newcommand{\gothn}{\mathfrak{n}}

\newcommand{\half}{\frac{1}{2}}
\newcommand{\be}{\begin{equation}}
\newcommand{\ee}{\end{equation}}
\newcommand\beq{\begin{eqnarray}}
\newcommand\eeq{\end{eqnarray}} 
\newcommand\eqn[1]{\label{eq:#1}} 
\newcommand\eq[1]{eq. (\ref{eq:#1})} 
\newcommand\Eq[1]{Eq. (\ref{eq:#1})}

\newcommand{\kfpm}{{k_{\mathrm F\pm}}}

\newcommand{\eV}{{\rm ~eV }}

\newcommand{\MeV}{{\rm ~MeV }}
\newcommand{\CA}{{\cal A}}

\newcommand{\CE}{{\cal E}}

\newcommand{\CL}{{\cal L}}

\newcommand{\nn}{\nonumber}

\newcommand{\mybar}[1]

\begin{document}

\preprint{INT-PUB-16-042}

\title{Energy Conservation and the Chiral Magnetic Effect}
\author{David B. Kaplan}
\email{dbkaplan@uw.edu}
\affiliation{%
Institute for Nuclear Theory, University of Washington, Seattle, WA 98195  
}%
\author{Sanjay Reddy}
\email{sareddy@uw.edu}
\affiliation{
Institute for Nuclear Theory, University of Washington, Seattle, WA 98195  
}%
\author{Srimoyee Sen}%
\email{srimoyee08@gmail.com}
\affiliation{
Department of Physics, University of Arizona, Tucson, AZ  85721
}%

\date{\today}

\begin{abstract}
We analyze the chiral magnetic effect in a homogeneous neutral plasma from the point of view of energy conservation, and construct an effective potential for the growth of maximally helical perturbations of the electromagnetic field. We show that a negative curvature at the origin of the potential, indicating instability of the plasma, is induced by a chiral asymmetry in electron Fermi energy, as opposed to number density, while the potential grows at large field value. It follows that the ground state for a plasma has zero magnetic helicity; a nonzero electron mass will allow an excited state of a plasma with nonzero helicity to relax to that ground state quickly.  We conclude that a chiral plasma instability triggered by weak interactions is not a viable mechanism for explaining magnetic fields in stars except possibly when dynamics drives the system far from equilibrium.
\end{abstract}


\maketitle

\section{Introduction}
There has been renewed interest in the question of how anomalies manifest on macroscopic scales. In particular, the role of chiral magnetic effect has been studied in the context of heavy-ion collisions, Weyl semi-metals, hot and dense plasma encountered in the early universe, and ultra dense matter encountered inside neutron stars and supernovae ~\cite{Kharzeev2014133,Akamatsu:2013pjd}. Here, the chiral magnetic effect refers to the phenomena in which a net chiral charge in the initial state triggers an exponential growth of electromagnetic fields.     Ever since the anomaly was first discussed in the context of its effect on matter \cite{Vilenkin:PhysRevD.22.3080} there has been speculation about whether parity violation in the weak interactions could be responsible for triggering such an instability.  Remarkably, it has been claimed  that a homogeneous and isotropic plasma with massless electrons is unstable,   with speculation that this  instability could lead to the generation of  long-scale helical magnetic fields in the early universe \cite{Boyarsky:PhysRevLett.109.111602}. It has also been claimed that this effect could explain the large magnetic field found in  some neutron stars (magnetars)  \cite{Dvornikov:2014uza,Dvornikov:2015ena,Sigl:2015xva}. A disconcerting feature of some these earlier studies is the lack of a clear discussion about how energy is conserved in such processes \footnote{In Ref.~\cite{Dvornikov:2015ena} claim to tap into the thermal energy 
of the star to generate the magnetic field. In this scenario it is unclear where entropy could be stored in the final state when the magnetic field saturates and the temperature approaches zero.}.

In this article we reexamine the chiral magnetic instability from the point of view of energy conservation and construct an effective potential governing the growth of perturbations, resolving along the way some confusions in the literature about the correct form of the current that appears in Maxwells equations. Similar approach
has been used in the past to derive the chiral magnetic current in \cite{Fukushima:2008xe, Nielsen:1983rb} in the absence of a background weak interaction. The result of our analysis is that the lowest energy configuration for a homogeneous and isotropic plasma has vanishing helical magnetic field and equal Fermi energies for left- and right-handed electrons. An idealized plasma with massless electrons with parity violating initial conditions can exhibit an instability called the chiral plasma instability and oscillate about a corresponding lowest energy state with nonzero magnetic helicity. These oscillations are damped by a finite electrical conductivity and an equilibrium state with non-vanishing magnetic helicity  and unequal Fermi energies for left- and right-handed electrons is reached. However, for a neutron star plasma with a finite electron mass, the chiral asymmetry in the initial state decays rapidly and the final state has no magnetic field or chiral asymmetry. Our analysis suggests that magnetic field amplification due to chiral plasma instability in neutron stars requires additional dynamics driving the plasma far from equilibrium.

\section{Chiral Magnetic Current}
The chiral magnetic instability is seen when solving Maxwell's equations in a plasma where there is a current proportional to the magnetic field; such a current  can arise when there is parity violation either in the underlying microphysics, or as an initial condition on the plasma. The relevant equation is
\beq
\vec \nabla\times \vec B - \frac{\partial\vec E}{\partial t} = \sigma \vec E + \xi \vec B\ ,
\eqn{max}\eeq
where the constitutive relation $\vec J =  \sigma \vec E + \xi \vec B$ ignores effects higher order in derivatives of the gauge field, such as a magnetic polarizability term $\nabla\times \vec B$, etc.  We consider the ansatz of a helical gauge field
\beq
A_0 = 0\ ,\qquad \vec A =\left(\hat x \,\cos kz - \hat y\,\sin k z \right) \CA_k(t)
\eqn{helical}\eeq
and with the assumption that $\xi$ is constant in time, one finds a maximally unstable mode at $k_\star = \xi/2$ growing exponentially in time as
\beq
\CA_{k_\star}(t) \sim \CA_{k_\star}(0) e^{t/\tau}\ ,\qquad  \tau =\frac{2}{\sqrt{\sigma^2 + \xi^2}-\sigma} \ .
\eqn{instab}
\eeq
As the $\xi \vec B$ contribution to the current can be derived from the effective action
\beq
\frac{\xi}{2}\int d^4x\, \epsilon_{ijk} A_i \partial_j A_k\ ,
\eeq
which is linear in derivatives, unlike the Maxwell term which is quadratic in derivatives,   it should be no surprise that  such a term could lead to instability and exponential growth at long wavelength.   This contribution to the current is related to the anomaly, but like the anomaly it can be computed by analyzing the effect of classical forces on fermions with a bottomless Dirac sea. That is the approach we will take here, paying close attention to energy conservation. Our analysis shows that the coefficient $\xi$ is not constant in time in physical systems of interest, such as neutron stars, and that in fact such exponential growth is not sustained.

\begin{figure}[t]
\centerline{\includegraphics[width=6 cm]{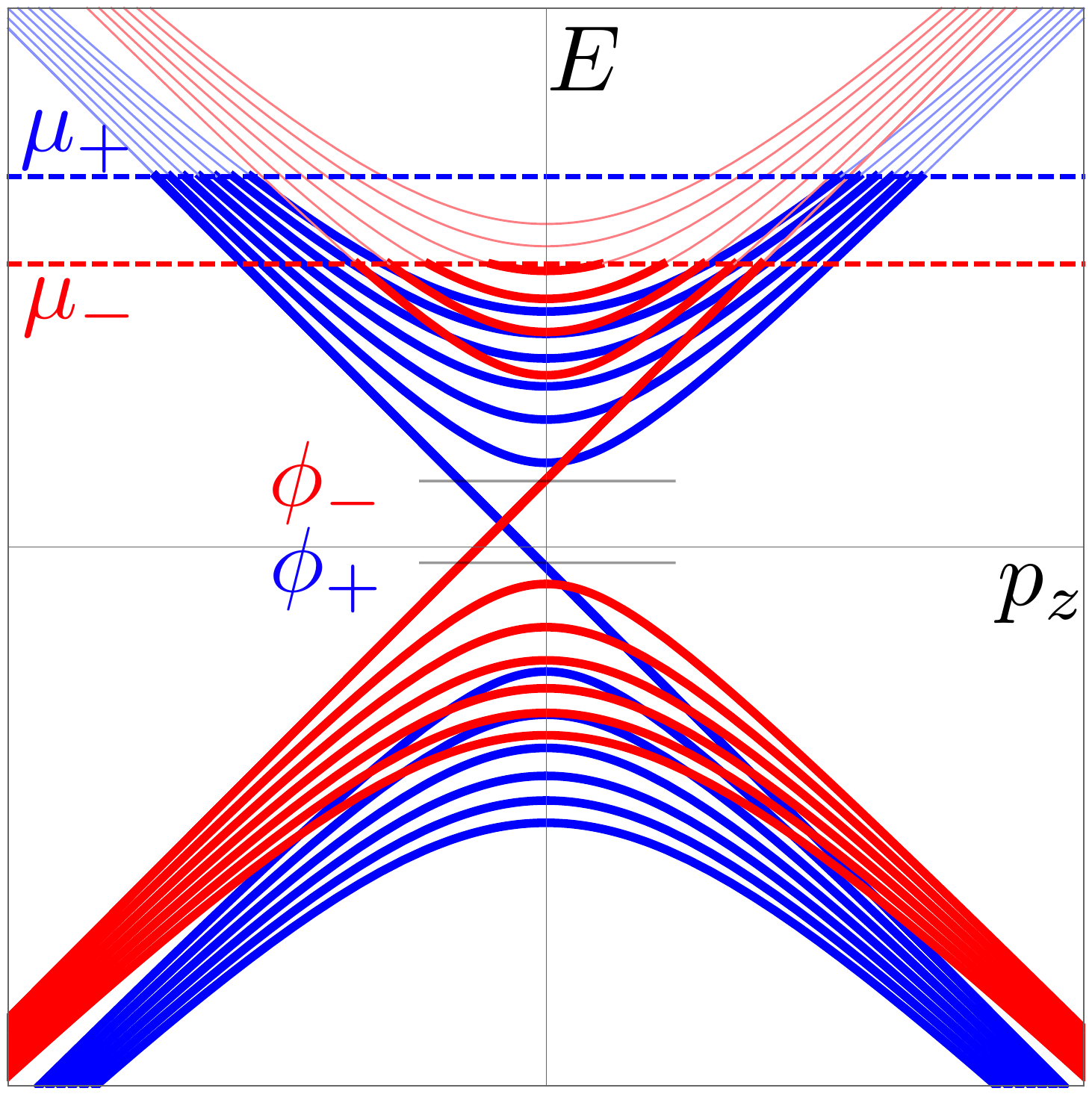}}
\caption{{\it The electron energy spectrum at zero temperature  in a magnetic field $\vec B=B \hat z$, plotted versus $p_z$ for positive (blue) and negative (red)  chirality states. The single particle energy levels are shifted vertically by $\phi_{\pm}$ respectively in the presence of a background potential due to neutral current interactions with the ambient nucleons.   Thick lines indicate occupied states, which are filled to Fermi energies $\mu_\pm$.
}}
\label{fig:spect}\end{figure}

We start by computing the chiral magnetic current for idealized electrons, protons and neutrons in a box in weak equilibrium at zero temperature with a constant magnetic field $B$ in the $z$ direction.     We take the electrons to be massless, and impose periodic boundary conditions.  In this case electron chirality is conserved and the right-handed ($\gamma_5=+1$) and left-handed ($\gamma_5=-1$) electrons will have independent chemical potentials denoted by $\mu_\pm$.  As usual, the chemical potentials are given by $\mu = \partial E/\partial N$, where $E$ is the energy and $N$ the particle number, and they equal the Fermi energy at  zero temperature.  We treat the neutral current interaction between the electrons and the nucleons in a mean field approximation, 
\beq
\CL_\text{NC} =-\phi_- \bar e\gamma_\mu P_L e - \phi_+ \bar e\gamma_\mu P_R e 
\eeq
where the mean fields are parametrically of order  $\phi_\pm \sim G_F n_n\simeq 5\eV (n_n/n_0)$, for a nuclear density of $n_0$ and
a nucleon density of $n_n$.   In the presence of magnetic field the electrons occupy Landau levels, where the energies for the electrons in the $n^{th}$ level are given by
\beq
\varepsilon^\pm_0  = \phi_\pm \mp p_z\ ,\qquad \varepsilon^\pm_n = \phi_\pm \pm \sqrt{p_z^2 +2 e Bn}\ ,\qquad n=1,2,\ldots\ .
\eeq
Here, for the lowest Landau level the $\mp$ in front of $p_z$  is determined by the chirality, while for the excited levels,  the $\pm$ in front of the square root occurs for both chiralities \footnote{The electron spin is anti-parallel to the magnetic field in the positive z-direction. Thus, electrons moving in the negative z-direction have positive helicity.}.  The electron states occupied within the box are then as depicted in Fig.~\ref{fig:spect}. The transverse density of states for each Landau level $g_n$ can be  derived by requiring that the total density of states revert to the free particle answer as the magnetic field is turned off, namely
\beq
 g_n  \xrightarrow[B\to 0]{} \frac{p_\perp d p_\perp}{2\pi}\ ,\qquad p_\perp^2\equiv \varepsilon_n^2 -p_z^2\ ,
\eeq
 with the result
 \beq
 g_n = \frac{e B}{2\pi}\ ,
 \eqn{gn} \eeq
 independent of $n$.

It is evident from Fig.~\ref{fig:spect} that this state has a nonzero electromagnetic current, and that it is proportional to $B$ since $g_n$ is proportional to $B$.  To compute what the current is we use energy conservation.  If we consider adding to the system an electric field parallel to the magnetic field, $\vec E = E\hat z$, then Maxwell's equations tell us that 
\beq
\frac{d}{dt} \CE_\text{em} \equiv \frac{d}{dt}\, \half\left(E^2+B^2\right) = - \vec E\cdot\vec J\ ,
\eeq
where $\CE_\text{em}$ is the energy density in the electromagnetic fields and $\vec J$ is the electromagnetic current.  We assume that we have arranged our plasma such that at time $t=0$ the electrons are in the zero temperature configuration shown in Fig.~\ref{fig:spect}, while the background protons which neutralize the system are fixed in place with constant density.  At time $t=dt$ the electrons will have picked up momentum
\beq
 d p_z = - e E dt
\eeq
and to linear order in $dt$ the change in the energy of the electrons is found to be
\beq
d\CE_{\rm e^-} =g_n \left(\mu_--\mu_+\right)\left(\frac{dp_z}{2\pi}\right) = \frac{2\alpha}{ \pi }\mu_5 E B \,dt\ ,
\eqn{dE}
\eeq
with the definition
\beq
\mu_5 \equiv \frac{\mu_+-\mu_-}{2}\ .
\eeq
This result is derived by noting that to order $dt$ the excited Landau levels do not contribute to a change in the total electron energy, since the contribution from electrons with positive and negative $p_z$ cancel each other. Thus the only contributions to $d\CE_{\rm e^-} $ come from the lowest Landau level, with the above result.  Since the total energy $\CE_{\rm tot}=(\CE_\text{em}+\CE_{\rm e^-} )$ must be conserved, the right side of the above equation must equal $ \vec E\cdot\vec J\ dt$, and it follows that the current in the $t=0$ configuration shown in Fig.~\ref{fig:spect} is
\beq
\vec J =\xi  \vec B\ ,\qquad \xi = \frac{2\alpha}{\pi}~\mu_5\ .
\eqn{Jval}
\eeq
Note that
the current is not proportional to the chiral number density in the lowest Landau level as claimed by \cite{Dvornikov:2014uza} and is instead proportional to the chiral chemical potential. While it may seem intuitive to associate the current with the difference between the number densities of electrons moving parallel and anti-parallel to $\vec{B}$, the subtlety associated with the anomaly which owes its origins to issues related to the UV regularization of the theory requires the existence of a current in the vacuum at the cut-off scale called the Chern-Simons current. When the Fermi energies of the L and R electrons are equal, this current exactly cancels the contribution to the current from occupied electron states in the lowest Landau level \cite{Landsteiner:2016led}. Our form for the chiral magnetic effect appears to differ from that in Ref.~ \cite{Landsteiner:2016led} only due to the unimportant convention that we define $\mu_5$ relative to the zero energy state of the perturbative QED vacuum ({\it i.e.} the vacuum outside a neutron star) while in  \cite{Landsteiner:2016led} $\mu_5$ is measured relative to the vertex of the Weylcone inside the medium.

The result in \eq{Jval} is independent of the assumption that the plasma be at zero temperature
and can be generalized to finite temperature as follows. Let $f_n^\chi(p,t)$ be the  distribution function for electrons with momentum $p \,\hat z$ in the $n^{th}$ Landau level with chirality $\chi$ at time $t$.  We assume that at a particular time $t$ this is given by the Fermi-Dirac distribution corresponding to temperature $ \beta^{-1}$ and chemical potential $\mu^\chi$, $f_n^\chi(p,t) = f_{\text{FD}}(\beta,  \varepsilon^\chi_{n}(p)-\mu^\chi)$, where $ \varepsilon^\chi_{n}(p)$ is the energy of an electron with chirality $\chi$ in the $n^{th}$ Landau level with momentum $p \hat z$. Then   the effect on the electrons of adiabatically applying an electric field for infinitesimal time $dt$ is to change the distribution function to 
\beq
f_n^\chi(p,t+dt)= f_{\text{FD}}(\beta, \varepsilon^\chi_{n}(p-dp)-\mu^\chi)  \ ,\qquad dp = -e E dt\ .
\eeq
This follows since each electron picks up momentum $dp$ due to the electric field, so that the number of electrons with momentum $p$ at time $t+dt$ is equal to the number of electrons that had had momentum $p-dp$ at time $t$.
Thus during this time interval $dt$ the electron energy density  has changed by 
\beq
d\CE_{\rm e^-}(t)&=&\sum_{\chi=\pm}\sum_n g_n\int_{-\infty}^\infty \frac{dp}{2\pi}\, \varepsilon^\chi_{n}(p)\left[ f_n^\chi( p,t+dt) - f_n^\chi(p,t)\right]\nn\\
&=&\frac{e B}{2\pi}\sum_{\chi=\pm}\sum_n  \int_{-\infty}^\infty \frac{dp}{2\pi} \, \varepsilon^\chi_{n}(p)\left[f_\text{FD}(\beta, \varepsilon^\chi_n(p-dp)-\mu^\chi) - f_\text{FD}(\beta, \varepsilon^\chi_{n}(p)-\mu^\chi\right] \nn\\
&=&\left(-e E dt \right)\frac{e B}{2\pi}\sum_{\chi=\pm}\sum_n \int_{-\infty}^\infty \frac{dp}{2\pi} \, \varepsilon^\chi_{n}(p) \left(-\frac{\partial}{\partial p}f_\text{FD}(\beta, \varepsilon^\chi_{n}(p)-\mu^\chi)\right)\ ,
\eeq
where we used $g_n = eB/2\pi$ from \eq{gn}. Since $ \varepsilon^\chi_{n}(p)$  is an even function of $p$ for the excited Landau levels, the contribution of these levels to the integrand is an odd function of $p$ and vanishes on integration, leaving us with only the contributions from the lowest Landau levels for each chirality. The linear dispersion relation for the lowest Landau level is 
\beq
   \varepsilon^\chi_{0}(p) = \begin{cases} -p + \phi_+ \ ,& \chi=+\cr  +p + \phi_-\ , & \chi = -\end{cases}\ ,
   \eeq
and   so we obtain
\beq
\frac{d\CE_{\rm e^-}}{dt} 
&=&\frac{\alpha EB}{\pi} \int_{-\infty}^\infty\, dp\,  \left[ (p + \phi_-) \frac{\partial}{\partial p} f_\text{FD}(\beta, p + \phi_- - \mu_-) + (-p + \phi_+) \frac{\partial}{\partial p} f_\text{FD}(\beta, -p + \phi_+ - \mu_+) \right]\cr&&\cr
&=& \frac{\alpha EB}{\pi} \int_{-\infty}^\infty\, dp\, \left[\left(p+\mu_-\right) - \left(p+ \mu_+\right)\right]\frac{\partial}{\partial p} f_\text{FD}(\beta, p)\cr&&\cr
&=& \frac{2\alpha EB}{\pi} \mu_5\ ,
\eqn{ftemp}
\eeq
where in the last line we used the fact that $f_\text{FD}(\beta,-\infty)=1$ and  $f_\text{FD}(\beta,+\infty)=0$, the characteristic of the Dirac sea which makes the anomaly temperature independent. The above result for $d\CE_{\rm e^-}/dt$  is the same as  obtained for zero temperature in \eq{dE}, and  so the current in \eq{Jval} is valid at all  temperatures.

Although the chiral potentials $\phi_\pm$ do not appear in the current, they affect the number densities which are given by
\beq
n_\pm = \frac{k^3_{\mathrm F\pm}}{6\pi^2}=\frac{(\mu_\pm-\phi_\pm)^3}{6\pi^2} \ ,
\eqn{densities}\eeq
where $\kfpm= (\mu_\pm-\phi_\pm)$ are Fermi momenta of the positive and negative chirality states, with corrections suppressed by  $O\left(\frac{e B}{\mu_\pm^2}\right)$, which assumes the Fermi  energies to be much greater than the Landau level splittings. 

The action of an electric field on electrons in the lowest Landau level 
changes not only  its kinetic energy, but also the potential energy associated with the parity violating 
electron-neutron interaction. The rate of change of this potential energy density is 
\beq
\frac{d\CE_{pot}}{dt} = \dot{n}_5 ~\phi_5\ \,,
\eeq
where the chiral charge density $ n_5$ and the chiral potential $\phi_5$ are given by  $ n_5=n_+ - n_-$ and $\phi_5=(\phi_+-\phi_-)/2$.  Had we neglected this contribution, we would have erroneously concluded that the current 
was given by  $\vec{J}= (2\alpha/\pi)~(\mu_5 - \phi_5) ~\vec{B}$ rather than the correct expression in \eq{Jval}.   

\section{The effective potential}

\subsection{General Analysis}
Having identified the correct current for the fermion configuration in Fig.~\ref{fig:spect} we now reconsider the dynamics of the plasma and show that the instability found in \eq{instab} is not relevant to the physical situations of interest unless a new source of energy is found to power the growth of helical fields, as well as a mechanism to convert the helical magnetic field at relatively short wavelength to a large-scale field.

We start by assuming that in the presence of an electric field the occupation numbers of the electrons can adjust quickly to that configuration through chirality conserving scattering processes, with only $\mu_\pm$ changing with time; this is equivalent to assuming the plasma is dissipationless with conductivity $\sigma=0$; we will relax this assumption later, but it should be clear that adding dissipative effects will not enhance instabilities. When $\xi$ is given by \eq{Jval} and $\sigma=0$, \eq{max} can be written as 
\beq
\ddot \CA_k(t)=-  \CA_k(t)  \Biggl(k  -\frac{2\alpha }{\pi}~\mu_5(t)\Biggr)k\,.
\eqn{feq}
\eeq
To derive an effective potential associated with the above equation of motion for ${\cal A}_k$ we need to relate the time dependent $\mu_5(t)$ to ${\cal A}_k(t)$. To do this we note that rapid scattering keeps the configuration in quasi-static equilibrium is chiral symmetry preserving for massless electrons, but chiral symmetry is violated in an electric field by electrons entering or leaving the Dirac sea, which can only happen in the lowest Landau level. The rate of change of the chiral densities is  given by
\beq
\dot n_\pm = \mp g_n \frac{\dot p_z}{2\pi} = \pm \frac{\alpha}{\pi}\vec E\cdot\vec B \ ,
\eqn{anom}\eeq
which reproduces the conventional anomaly calculation (see, for example, the discussion in Ref.~\cite{Kaplan:2009yg}). If we use the helical ansatz for the gauge field of \eq{helical} then
\beq
 \frac{\alpha}{\pi}\vec E\cdot\vec B = -\frac{\alpha k}{2\pi} \frac{d}{dt} \CA_k^2\ ,
 \eeq
 and so the anomaly equation \eq{anom} integrates to give us two constants of motion $ {\gothn}_\pm$ given by 
  \beq
  {\gothn}_\pm = n_\pm(t) \pm \frac{\alpha k}{ 2\pi} \CA_k(t)^2  \ .
\eqn{intanom}
 \eeq

Combining the above result with the simplifying assumption that the Landau level splittings are small compared to the electron Fermi energies, $\mu^2_\pm \gg eB$, we can use the expression in \eq{densities} for the number densities $n_\pm$ of electrons and solve for the chemical potentials $\mu_\pm(t)$   in terms  $\CA_k(t)$ and the constants $\gothn_\pm$,
 \beq
\mu_\pm(t) = \phi_\pm + k_{F_\pm}(t)
\eqn{musol}\eeq
where the fermi momenta of the positive and negative helicity states, 
\beq
k_{F_\pm}(t)= \left[6\pi^2\left(\gothn_\pm \mp  \frac{\alpha k}{2\pi}\CA_k(t)^2\right) \right]^{1/3} \,, 
\eeq
are obtained by using \eq{densities}, where the real cube root is implied.

We can now substitute these results for $\mu_5(t) =( \mu_+(t) - \mu_-(t))/2$ in \eq{feq} to obtain
\beq
\ddot \CA_k(t)  =-  \CA_k(t)\left[k  - \frac{2\alpha }{\pi }\left( \phi_5 + \frac{k_{F_+}(t)-k_{F_-}(t)}{2})\right)\right] k
\eqn{feq2}
\eeq
where $\phi_5 = (\phi_+-\phi_-)/2$ is the mean field coupling to the axial electron current. The above equation is integrable, allowing us to write
\beq
\frac{\dot \CA_k^2}{2} + V(\CA_k)= \CE_{\rm tot}\ ,
\eqn{etot}
\eeq
analogous to a particle with unit mass moving  in a potential $V$ with a conserved total energy $ \CE_{\rm tot}$, where
\beq
V(\CA_k)= \frac{ \CA_k^2 k^2 }{2} - \frac{\alpha \CA_k^2 k \phi_5}{\pi}   + \frac{ k^4_{F_+}(t) +  k^4_{F_-}(t)}{8\pi^2}   \,.
\eqn{potential}
\eeq

If we assume  $\CA_k=0$ at initial time $t=0$ and expand $V$ for small $\CA_k$ we find 
\beq
V(\CA_k) = V(0) + \half \CA_k^2  \left[(k -k_\star)^2 - k_\star^2\right] + O(\CA_k^4)\,,
\eqn{curve}\eeq
with 
\beq
k_\star = -\frac{\alpha \mu_5(0)}{\pi}\,,
\eqn{kstar}\eeq
where $\mu_5(0)$ is the chiral chemical potential at the initial time $t=0$. The potential $V(\CA_k)$ exhibits an instability for $k_\star\ne 0$ maximized for wavenumber $k=k_\star$, and \eq{kstar}  shows explicitly that the instability is triggered by a difference in the initial values for the Fermi energies $\mu_\pm$, and not by the neutral current mean field $\phi_5$.  On the other hand,  for large field amplitude $\CA_k$ the potential $V$ behaves as
\begin{figure}[t]
\centerline{\includegraphics[width=6 cm]{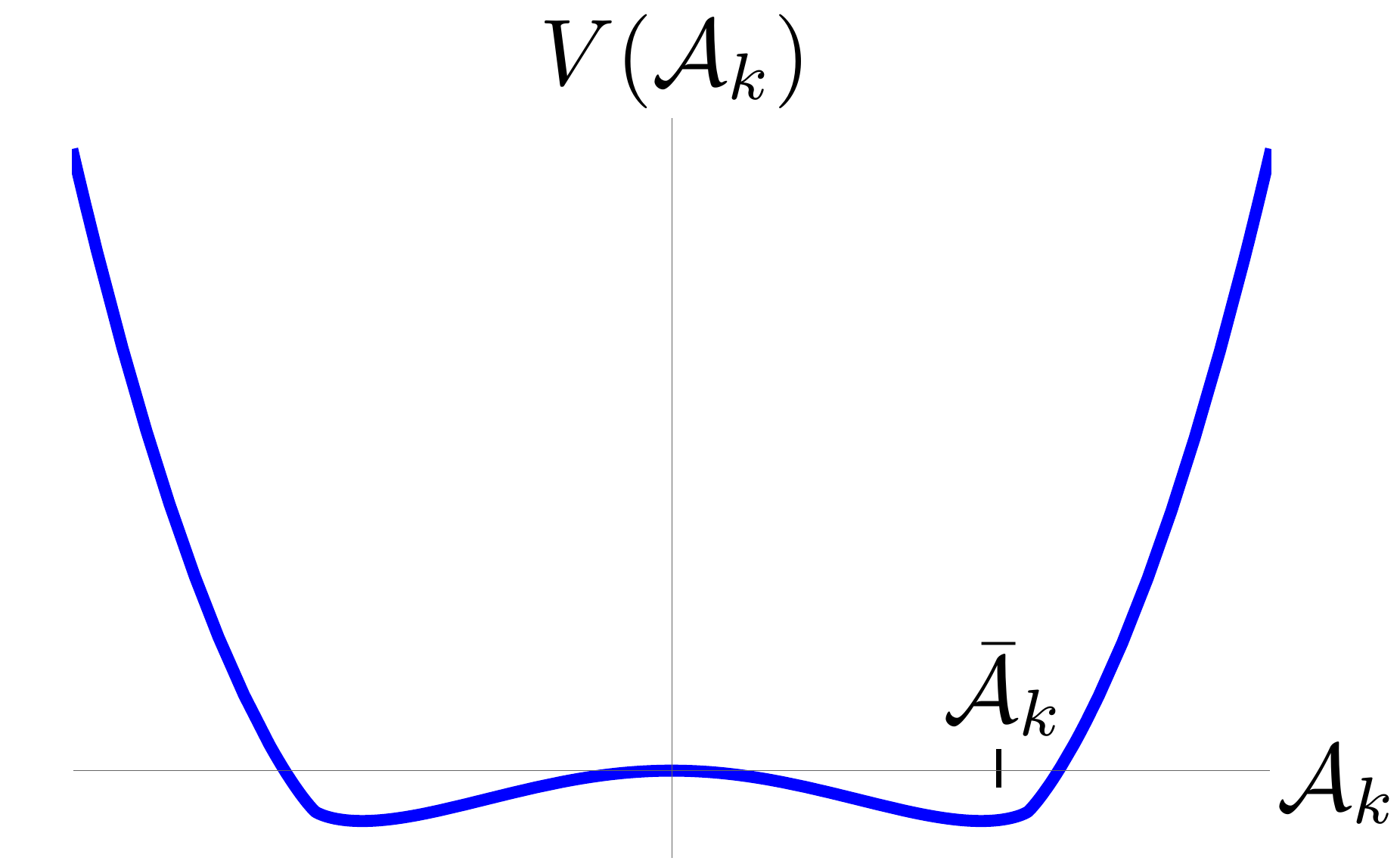}}
\caption{{\it The potential $V(\CA_k)$ in \eq{potential}  governing the time evolution of $\CA_k$ for a particular set of initial conditions and $k=k_\star$, defined  in \eq{curve}.}}
\label{fig:pot0}\end{figure}
\beq
V(\CA_k) \xrightarrow[ |\CA_k|\to\infty]{} \frac{\pi^2}{4}\left\vert  \frac{3k \alpha}{\pi^2}\CA_k^2\right\vert^{4/3}
\eeq
and so we see that there are no runaway solutions, despite claims to the contrary in the literature.  This is the effect of energy conservation. The generic shape of the potential $V$ is shown in Fig.~\ref{fig:pot0}.

By using \eq{musol}, we can rewrite \eq{etot} as 
\beq 
\left(\frac{E^2+B^2}{2}\right)+ \frac{ k^4_{F_+}(t) +  k^4_{F_-}(t)}{8\pi^2}+ \phi_5 \frac{2\alpha}{\pi}\int^t \vec E\cdot \vec B=\CE_{\rm tot}\,, 
\eqn{econserve}
\eeq
where the first term on the left side of the above equation is the energy density of the electromagnetic field,  the second term is the energy density of the electron gas, and the third term is the interaction energy of  the electron axial charge density with the background neutral current potential $\phi_5$, equal to $n_5 \phi_5$, up to an overall integration constant.  
Here, we have used the fact that the electric and magnetic fields associated with the helical field in \eq{helical} are given by $E^2=\dot \CA_k^2$,  $B^2=k^2 \CA_k^2  $, and  $\vec E\cdot \vec B = -k \CA_k \dot \CA_k$. \Eq{econserve} clarifies the meaning of the different terms appearing in \eq{etot} and \eq{potential}.

\subsection{Expansion for small chiral asymmetry}

Since in practice chiral asymmetries and weak interaction mean fields will usually be very small, we treat $\mu_5$, $\phi_5$,  $\gothn_5$ and $k$ as $O(\epsilon)$, and redo the above analysis to $O(\epsilon^2)$.  In this case one finds the potential 
\beq
V(\CA_k) =  V(0) + \half \CA_k^2  \left[(k - k_\star)^2 -  k_\star^2\right]  + \CA_k^4 \frac{\alpha^2 k^2}{2\mu^2}\ ,
\eeq
which for $(k-k_\star)^2 < k_\star^2$ is the classic double-well potential with minimum for $k=k_\star$ at
\beq
\bar \CA_{k_\star} = \frac{\mu}{\sqrt{2}\alpha}\ .
\eqn{aks}\eeq 
It is somewhat remarkable that the gauge potential at the minimum is independent of the initial chiral asymmetry; however, in order  to see the system relax to such a state one needs to include dissipation through nonzero conductivity.  

\section{Time Evolution}
\subsection{Finite conductivity}
\label{dissipation} 
From our above discussion we see that in the absence of dissipation, an initial chiral imbalance in the system will trigger an instability in the gauge field which will subsequently oscillate in the effective potential. With nonzero conductivity, a damping term $\sigma \dot \CA_k$ appears on the left side of \eq{feq} which is now given by 
\beq
\ddot\CA_k + \sigma \dot\CA_k&=&-\CA_k\left(k-\frac{2 \alpha}{\pi}\mu_5(t)\right)k\,.
\eqn{fsigma}
\eeq
For $k=k_\star$ the dissipative term allows the field to decay to the stationary ground state where $\CA_k=\bar \CA_{k_\star}$ and $\dot \CA_k=0$ on a time scale given approximately by
\beq
\tau_{k_\star} =\frac{\pi \sigma }{2 \alpha \mu_5(0) k_\star} =\frac{\sigma}{2 k^2_\star} \ ,
\eqn{taustar}\eeq
 where 
\beq
\sigma \simeq \sigma_0~\left(\frac{\mu}{100~\MeV }\right)^{9/2}~\left(\frac{10^8~{\rm K}}{T}\right)^{2}\, 
\eeq 
is the electrical conductivity of degenerate matter with $\sigma_0\approx 10^{29}$ s$^{-1}$ \cite{BaymPethickPines:1969b}. Given \eq{kstar} this allows us to express $\tau_{k_\star}$ as
\beq
\tau_{k_\star} 
\simeq  10^{-12}~{\rm sec}~\left(\frac{\mu}{\mu_5(0)}\right)^2\left(\frac{\mu}{100~\MeV }\right)^{5/2}~\left(\frac{10^8~{\rm K}}{T}\right)^{2}
 \eqn{tks}\eeq

Once the system has reached this equilibrium state at late  time $\tau > \tau_{k_\star}$,   the chiral chemical potential (for $k=k_\star$) has adjusted to half its original  value   
\beq
\mu_5(t)\to \frac{\mu_5(0)}{2}\ ,
\eeq
and the modes of magnetic field given by $\vec{B}_k = k \bar{\CA}_k \left(\hat x \,\cos kz - \hat y\,\sin k z \right) $ satisfy   
\beq 
\bar B^2_k=\begin{cases} \frac{\mu^2 }{2  \alpha^2} \left(k_\star^2 -(k-k_\star)^2\right)  &  \text{for }k_\star^2>(k-k_\star)^2\cr 0 &\text{otherwise}\end{cases}\ .
\eeq 
At the optimal wave number $k=k_\star$  we have
\beq 
\bar B^2_{k_\star}=\frac{\mu^2 k_\star^2}{2  \alpha^2}= \frac{\mu^2\mu^2_5(0)}{2 \pi^2}\,.
\eeq 
It is instructive to rearrange the above equation to yield the wavelength of the magnetic field as a function of its magnitude, with the result
\beq
\lambda_\star =\frac{2\pi}{k_\star}= \frac{2\pi \mu}{\sqrt{2} \alpha \bar{B}_{k_\star}} =\left(6.2\times 10^{-7} \, {\rm m}\right)\times \left(\frac{\mu}{100\MeV}\right)\left(\frac{10^{12}\text{ gauss}}{\bar B_{k_\star}}\right) ,
\eeq
indicating that this mechanism, assuming growth of a single mode at the maximally unstable wavenumber $k_\star$, cannot produce large scale strong magnetic fields on astrophysical scales, where a magnetar can exhibit fields as large as $10^{15}$ gauss on a length scale of kilometers. A second mechanism is needed to convert the short wavelength modes into large scale fields. However, as we shall discuss below, any such conversion must occur rapidly or it will fail to grow due to chiral symmetry violation from the nonzero electron mass.

\subsection{Finite electron mass}
\label{mass}
A nonzero electron mass  violates chirality and introduces into the problem a new time scale $\tau_{m}$, the time scale for an electron to change chirality. We shall now consider the time evolution with both $m_e\ne 0$ and $\sigma\ne 0$ of an initial excited state with  $\mu_5(0)\ne 0$ and  $\vec{E}(0)=\vec{B}(0)=0$, or equivalently, $\dot\CA_k(0)=\CA_k(0)=0$. This particular initial condition was chosen  to demonstrate that a nonzero initial chiral chemical potential caused either due to the background weak interaction or due to any other non-equilibrium process decays rapidly even at the extreme densities realized inside neutron stars, where electrons are ultra-relativistic with $\mu \gtrsim  200 ~m_e$.  

In this case, it is convenient to write evolution equations in terms of a time dependent chiral chemical potential for the electrons as opposed to a time dependent chiral charge density. When $m_e\neq 0$ electron chirality can change either due to Rutherford scattering off protons, or due to Compton scattering. The chirality flipping rate due to these reactions needs to be included in the equation for the time evolution of $\mu_5$. In earlier work in Ref.~\cite{Grabowska:2014efa} it was found that Rutherford scattering dominates in dense plasmas encountered in neutron stars. When protons are degenerate the relaxation time $\tau_{\rm f}$ was calculated in Ref.~\cite{Dvornikov:2015iua}, 
\beq
&& \tau_{m} 
= \frac{1}{\alpha^2 \kappa}  \left(\frac{\mu}{m_e^2}\right) \left(\frac{E_{\rm Fp}}{T}\right) \simeq 10^{-12}~\text{sec} \left(\frac{\mu}{100\MeV}\right)^3 \left(\frac{T}{10^8\,{\rm K}}\right)
\eqn{masseq3}
\eeq
where $\kappa \approx 3 $ is a numerical factor than can receive modest corrections from strong interactions, and $E_{\rm Fp}\simeq \mu^2/2M$ is the proton Fermi energy and $M$ is the proton mass. Similarly the rate of change of chiral chemical potential
due to the chiral magnetic effect alone can be easily obtained in the $\mu_5 \ll \mu$ limit by taking a time derivative of ~\eq{intanom} and setting the time derivative of $\gothn_{\pm}=0$. Combining the chirality changing rates obtained from ~\eq{anom} due to the anomaly, and  ~\eq{masseq3} due to the mass term, we have
\beq
\dot{\mu}_5(t)&=&-\frac{\mu_5(t)}{\tau_{m}}-\frac{2\pi \alpha}{\mu^2}k \CA_k\dot\CA_k\qquad (\mu_5\ll\mu)\ .
\eqn{masseq4}
\eeq
We need to solve the above equation along with \eq{fsigma} to obtain the time evolution of $\mu_5(t)$ and $\CA_k(t)$. 

We see that with the finite conductivity wanting  $\CA_k$  to settle down to the time independent value $\bar\CA_k$ in time $\sim \tau_{k_\star}$, the nonzero electron mass tries to drive $\mu_5$ to zero on the time scale $\tau_{m}$ and consequently neither a chiral chemical potential nor magnetic fields can persist on longer timescales. The transient evolution on shorter timescales will depend on the magnitude of the initial chiral imbalance $\mu_5(0)$. 
When $\mu_5(0) \ll \mu$, it follows from \eq{tks} that $ \tau_{m} \ll \tau_{k_\star} $;  the chiral chemical potential $\mu_5(t)$ and the associated chiral magnetic current will decay exponentially on the timescale $\tau_{m}$  and vanish before it can source the growth of a helical magnetic field (recall that the value of $k$  in \eq{masseq4} for unstable modes is proportional to $\mu_5$). 

 In the opposite limit, when $ \tau_{k_\star} \ll \tau_{m} $, which is realized only for extremely large $\mu_5(0)$ of order $\mu$, and \eq{masseq4} is no longer valid.  However, it is still easy to characterize  the evolution: on the shorter timescale $\tau_{k_\star}$, the system will reach the  equilibrium gauge field configuration $\bar \CA_{k_\star}$ given  in \eq{aks}; subsequently, on a timescale $\tau_{m}$ the violation of chiral symmetry equilibrates the Fermi energies of the two chiralities and drives $\mu_5$ to zero; this in turn eliminates the chiral magnetic current of \eq{Jval} which is required to support a nonzero magnetic field. However, if an inverse cascade can convert the short wavelength helical field into long wavelength poloidal or toroidal fields, which are sustained by conventional currents in the highly conducting neutron star plasma, a strong large scale field can emerge and be stable on astrophysical timescales
 \cite{BaymPethickPines:1969b}.  Earlier work has shown that when a spectrum of magnetic modes is included, an inverse cascade converts short wavelength helical magnetic modes into large scale magnetic fields \cite{Boyarsky:2012, Tashiro:2012mf,Pavlovic:2016gac,Buividovich:2015jfa}. A similar analysis which takes into account the rapid decay of $\mu_5$ due to the mass term is needed to determine if this conversion can occur on short time scales. This would involve a numerical study and is beyond the scope of this work.
 
\section{Conclusions}
We derived a general expression for the effective potential for helical electromagnetic field fluctuations in the presence of a general chiral imbalance in the electrons, as well as chiral mean fields arising from the parity violating electron-nucleon interaction. This simple derivation allowed us to establish that a plasma with an initial chiral imbalance in the electron Fermi energy (or nonzero chiral chemical potential) will relax to an equilibrium state with no net magnetic field or chiral chemical potential. We found that the evolution to equilibrium has two distinct timescales. For small initial chiral imbalance $\mu_5(0)\ll \mu$ with the electron mass induced chirality flipping transitions damp the chiral magnetic current on a timescale faster than is necessary to generate a helical magnetic field. In the extreme case when the imbalance is large with $\mu_5(0)\sim  \mu$, a transient short wavelength helical field is generated but it too decays on the timescale $\tau_{\rm m}$ associated with chirality flipping transitions, unless an inverse cascade is able to redistribute magnetic energy on larger scales on a shorter timescale. Although earlier work has shown that  an inverse cascade drives the helical magnetic field to larger wavelengths  \cite{Boyarsky:2012, Tashiro:2012mf,Pavlovic:2016gac,Buividovich:2015jfa}, it is essential that this large scale field be supported by  conventional currents which persist when $\mu_5=0$. If such a process operates in neutron stars, a large value of $\mu_5$ in the initial state may suffice to develop large-scale fields of astrophysical relevance.
   
Our analysis shows that the chiral magnetic current in a plasma, being proportional to $\mu_5$, cannot be sustained on timescales larger than $\tau_m$, even in the presence of the parity violating electron-neutron interaction, without a source of energy to sustain the chiral asymmetry, given its rapid violation by the electron mass. In Ref.~\cite{Sigl:2015xva} the authors suggest that turbulence can be an energy source that drives weak interactions out of equilibrium and sustains a chiral chemical potential on longer timescales. Here, a seed magnetic field of modest strength ensures that even relatively small perturbations of $\mu_5$ can produce the current necessary to drive the instability.     
They find that a modest amplification of magnetic field strength is possible under some conditions realized during the proto-neutron star phase subsequent to the supernova explosion.  This possibility warrants further study where effects due to the inverse cascade, spatial gradients of the chiral chemical potential \cite{Chen:2016xtg,Gorbar:2016qfh}, and other aspects essential to modeling magnetic field evolution in the neutron stars are included.   

\section{Acknowledgements}
We thank one of our referees for pointing out sign errors that existed in an earlier version of this work.
DK was supported in part by DOE Grant No. DE-FG02-00ER41132 and by the Thomas L. and Margo G. Wyckoff Endowed Faculty Fellowship. SR was supported by DOE Grant No. DE-FG02-00ER41132 and SS was supported DOE Grant No. DE-FG02-04ER41338. 

\bibliographystyle{apsrev4-1}
\bibliography{electronneutron}

\end{document}